\documentclass[aps,prd,twocolumn,eqsecnum,showpacs,amsmath,nofootinbib]{revtex4-1}
\usepackage{color,graphicx}
\usepackage{amsfonts}
\usepackage{amssymb}

\begin{document}

\title{Gravitational lensing in Tangherlini spacetime in the weak gravitational field and the strong gravitational field}

\author{Naoki Tsukamoto${}^{1}$}\email{tsukamoto@fudan.edu.cn}
\author{Takao Kitamura${}^{2}$} 
\author{Koki Nakajima${}^{2}$}
\author{Hideki Asada${}^{2}$} 
\affiliation{
${}^{1}$Center for Field Theory and Particle Physics \& Department of Physics, Fudan University, Shanghai 200433, China\\
${}^{2}$Faculty of Science and Technology, Hirosaki University, Hirosaki 036-8561, Japan}
\date{\today}

\begin{abstract}
The gravitational lensing effects in the weak gravitational field by exotic lenses have been investigated intensively 
to find nonluminous exotic objects. 
Gravitational lensing based on $1/r^n$ fall-off metric,
as a one-parameter model that can treat by hand both
the Schwarzschild lens (n=1) and the Ellis wormhole (n=2) in the weak field,
has been recently studied.
Only for $n=1$ case, however, it has been explicitly shown that effects of relativistic
lens images by the strong field on the light curve can be neglected. 
We discuss whether relativistic images by the strong field can be neglected for $n>1$ in the 
Tangherlini spacetime which is one of the simplest models for our purpose.
We calculate the divergent part of the deflection angle for arbitrary $n$
and the regular part for $n=1$, $2$ and $4$ in the strong field limit, 
the deflection angle for arbitrary $n$ under the weak gravitational approximation. 
We also compare the radius of the Einstein ring with the radii of the relativistic Einstein rings for arbitrary $n$.
We conclude that the images in the strong gravitational field have little effect on the total light curve
and that the time-symmetric demagnification
parts in the light curve will appear even after taking account of the images in the strong gravitational field for $n>1$.
\end{abstract}



\maketitle

\section{Introduction}
Gravitational lensing is useful to survey nonluminous objects for example extrasolar planets and dark matters 
(see~\cite{Schneider_Ehlers_Falco_1992,Perlick_2004_Living_Rev,Perlick_2010,Petters_Levine_Wambsganss_2001, 
Schneider_Kochanek_Wambsganss_2006,Wambsganss_1998,Bartelmann_2010}
for the details of the gravitational lensing and references therein). 
The gravitational lensing effects of the Schwarzschild lens in the weak gravitational field have been investigated for a hundred years.

Exotic objects such as wormholes, cosmic strings also cause gravitational lensing effects.
The gravitational lenses in wormhole spacetimes were pioneered 
by Kim and Cho~\cite{Kim_Cho_1994} and Cramer~\textit{et al.}~\cite{Cramer_Forward_Morris_Visser_Benford_Landis_1995}   
and then the gravitational lensing effects in various wormhole spacetimes have been investigated~\cite{Safonova_Torres_Romero_2001_Jan,
Eiroa_Romero_Torres_2001,Safonova_Torres_Romero_2001_Nov,Safonova_Torres_2002,Nandi_Zhang_Zakharov_2006,Rahaman_Kalam_Chakraborty_2007,
Dey_Sen_2008,Takahashi_Asada_2013,Kuhfittig:2013hva} (See Visser~\cite{Visser_1995} for the detail of wormholes).

The Ellis wormhole~\cite{Ellis_1973,Bronnikov_1973} is the simplest and earliest wormhole 
of the Morris-Thorne class~\cite{Morris_Thorne_1988,Morris_Thorne_Yurtsever_1988}.
Recently, Bronnikov~\textit{et al.} have shown that the Ellis wormhole metric supported by a perfect fluid with negative density
and a source-free radial electric or magnetic field 
is linearly stable under the spherically symmetric perturbations and axial perturbations~\cite{Bronnikov:2013coa}
while the Ellis wormhole with the same metric supported by a phantom scalar field is unstable~\cite{Armendariz-Picon_2002,
Shinkai_Hayward_2002,Gonzalez_Guzman_Sarbach_2008_I,Gonzalez_Guzman_Sarbach_2008_II,
Doroshkevich_Hansen_Novikov_Shatskiy_2009,Bronnikov:2011if}.
The former is the first example of a stable wormhole which is not supported by thin shells in general relativity.
The gravitational lensing in the Ellis wormhole spacetime has been investigated intensively~\cite{Chetouani_Clement_1984,Perlick_2004_Phys_Rev_D,
Nandi_Zhang_Zakharov_2006,Dey_Sen_2008,Abe_2010,Bhattacharya_Potapov_2010,Toki_Kitamura_Asada_Abe_2011,Nakajima_Asada_2012,Gibbons_Vyska_2012,
Tejeiro_Larranaga_2012,Tsukamoto_Harada_Yajima_2012,Tsukamoto_Harada_2013,Yoo_Harada_Tsukamoto_2013,Takahashi_Asada_2013} 
because of its simplicity and its features which 
are caused by the gravitational potential which decreases as the inverse square of a radial coordinate $r$ under the weak-field approximation.
Abe found numerically that the light curves of the microlensing by the Ellis wormhole show the time-symmetric demagnification 
under the weak-field approximation~\cite{Abe_2010}. 
This means that 
the theorem that total magnification is always larger than the unity is true 
for mass lenses with the gravitational potential which has the form of $1/r$
but it is not true for exotic lens objects with the gravitational potential which has the form of $1/r^{2}$ under the weak-field approximation.

Gravitational lensing based on $1/r^n$ fall-off metric,
as a one-parameter model that can treat by hand both
the Schwarzschild lens $(n=1)$ and the Ellis wormhole $(n=2)$ in the weak field,
has been recently studied~\cite{Tsukamoto_Harada_2013,Kitamura_Nakajima_Asada_2013,
Izumi_Hagiwara_Nakajima_Kitamura_Asada_2013,Kitamura_Izumi_Nakajima_Hagiwara_Asada_2013,Nakajima:2014nba}.
In particular, Kitamura~\textit{et al.} showed that the demagnification of the light curves appears in the case $n>1$~\cite{Kitamura_Nakajima_Asada_2013}.
This implies that the surveys of microlensing events are a good way 
to search exotic objects including the Ellis wormhole in our Galaxy and so on.
However, these studies concentrate on the weak gravitational field 
and do not cover the gravitational lensing effects in the strong gravitational field.

The timelike and null geodesics in the Schwarzschild spacetime 
were studied by Hagihara~\cite{Hagihara_1931} and Darwin~\cite{Darwin_1959,Darwin_1961}.
Darwin pointed out the existence of the relativistic images which are a series of faint images 
lying just outside the photon sphere~\cite{Darwin_1959} in the Schwarzschild spacetime.
A countably infinite number of relativistic images are
formed in a spherically symmetric static spacetime,
if the radius of the lens object is smaller than a photon sphere
($=3r_{g}/2$, where $r_{g}$ is the radius of the event horizon in the Schwarzschild spacetime)~\cite{Bozza_2002,Hasse_Perlick_2002,Perlick_2004_Phys_Rev_D}.
The gravitational lensing in the strong gravitational field by various black holes and wormholes 
has been investigated intensively in the recent decade 
(see~\cite{Virbhadra_Ellis_2000,Virbhadra_Keeton_2008,Virbhadra_2009,Bozza_2010,Bozza_Mancini_2012,AzregAinou:2012xv} and references therein). 
In a series of the papers on the relativistic images,
the observables for the suppermassive object at the center of our Galaxy
with instruments which can measure the relativistic images are discussed intensively.

The Tangherlini solution is an exact solution of the Einstein equation
in every dimension, where the solution is spherically symmetric, static and
asymptotically flat. Here, as a generalization of the Einstein-Hilbert action
to every dimension, the action for gravity is assumed to be the Ricci scalar
even in $d \neq 4$ dimensions for its simplicity, though the physics behind
this assumption in higher dimensions is not clear. Then, the variation of
the gravity action with respect to the metric provides the vacuum field
equation as "Ricci = 0". It follows that the Tangherlini solution recovers 
the Schwarzschild solution in four dimensions.

An exact form of the metric with $1/r^n$ in the weak field limit would
enable us to investigate such exotic lenses not only in the weak field
but also in the strong field. 
Only for $n=1$ case, 
it has been explicitly shown that effects of relativistic
lens images by the strong field on light curves can be neglected~\cite{Bozza_Capozziello_Iovane_Scarpetta_2001}. 
For $n > 1$, this issue
has not been addressed yet. Therefore, the main purpose of the present paper
is to discuss whether relativistic images by the strong field can be neglected
for $n > 1$. It is likely that a spacetime geometry with $1/r^n$
fall-off in the weak field is not unique but has many variants.
The Tangherlini solution is one of the simplest models for our purpose.
For recent investigations of exotic gravitational lenses, therefore,
it would be of physical interest to examine the Tangherlini lens both
in the weak field and in the strong one.

The gravitational lens in the strong field limit is related to the other phenomena 
such as the quasinormal modes of a black hole \cite{Stefanov:2010xz,Wei:2013mda}
and the high-energy absorption cross section \cite{Wei:2011zw}
which are caused by the nature of the null geodesic near the photon sphere. 
Thus, the investigation of gravitational lensing effects of the all-dimensional black hole 
in the strong field limit would give us a new perspective on the all-dimensional black hole.

This paper is organized as follows.
In Sec.~II, we review the null geodesic in the Tangherlini spacetime and investigate the deflection angle of light.
In Secs.~III and IV, we will investigate the deflection angles in the weak field approximation and in the strong field limit.
In Sec.~V, we study the gravitational lens effects in the strong field limit in the Tangherlini spacetime.
In Sec.~VI, we summarize our results.
In this paper we use the units in which the light speed $c=1$ and Newton's constant $G=1$.

\section{Deflection angle of light in Tangherlini spacetime}
The line element of the Tangherlini spacetime is given by \cite{Tangherlini_1963}
\begin{eqnarray}\label{eq:Tangherlini}
ds^2
&=&-\left[ 1-\left( \frac{r_{g}}{r} \right)^{d-3} \right]dt^2 \nonumber\\
&&+\frac{dr^2}{1-\left( \frac{r_{g}}{r} \right)^{d-3} } +r^{2}d\sigma^{2}_{d-2},
\end{eqnarray}
where $r_{g}$ is the event horizon radius and $d\sigma^{2}_{d-2}$ is the line element on the unit $(d-2)$-sphere which is given by    
\begin{eqnarray}
d\sigma^{2}_{d-2}=d\theta_{1}^{2}+\sum^{d-3}_{j=2} \prod^{j-1}_{i=1} \sin^{2}\theta_{i}d\theta_{j}^{2} +\prod^{d-3}_{i=1} \sin^{2}\theta_{i}d\phi^{2},
\end{eqnarray}
where $\theta_{i} \in [0,\pi]$ and $\phi \in [0,2\pi]$ are angles on the $(d-2)$-sphere and the integer $i$ runs from $1$ into $d-3$.
The event horizon radius $r_{g}$ is given by
\begin{eqnarray}
r_{g}=\frac{16\pi M}{(d-2)A_{d-2}},
\end{eqnarray}
where $M$ is the mass and $A_{d-2}$ is the area of the unit $(d-2)$-sphere which is given by
\begin{eqnarray}
A_{d-2}=\frac{2\pi^{\frac{d-1}{2}}}{\Gamma \left( \frac{d-1}{2} \right)}.
\end{eqnarray}

Because of the stationarity and axisymmetry, we can define the conserved energy of a photon as $E\equiv -g_{\mu\nu}t^{\mu}k^{\nu}$ 
and the conserved angular momentum of a photon as $L\equiv g_{\mu\nu}\phi^{\mu}k^{\nu}$, respectively.
Here, $t^{\mu}\partial_{\mu}=\partial_{t}$, $\phi^{\mu}\partial_{\mu}=\partial_{\phi}$ and $k^{\mu}$ 
correspond to the time translational Killing vector, the axial Killing vector and the photon wave number, respectively.
We assume that the conserved energy $E$ is positive. 

For its simplicity, we can set
$\sin\theta_{i}=1$ and consider the induced line element
\begin{eqnarray}\label{eq:Tangherlini}
ds^2=-\left[ 1-\left( \frac{r_{g}}{r} \right)^{n} \right]dt^2 +\frac{dr^2}{1-\left( \frac{r_{g}}{r} \right)^{n} } +r^{2}d\phi^{2},
\end{eqnarray}
where $n\equiv d-3$.
Note that our four-dimensional world may be off-center in the Tangherlini solution.
However, it is unlikely that such an off-center configuration is stable.
Therefore, we consider a four-dimensional section through the center of the solution. 
From $k^{\mu}k_{\mu}=0$,
the equation of the photon trajectory is obtained as
\begin{eqnarray}\label{eq:Photon_trajectory}
\left( \frac{dr}{d\phi} \right)^2
=r^{4}G(r,b),
\end{eqnarray}
where
\begin{eqnarray}\label{eq:G_definition}
G(r,b) \equiv \frac{1}{b^{2}}-\frac{1}{r^{2}}+ \frac{r_{g}^{n}}{r^{n+2}},
\end{eqnarray}
and $b \equiv L/E$ is the impact parameter of the photon.
We can concentrate ourselves on $L>0$ and $b>0$ because of spherical symmetry.

The equation $G(r,b)=0$ has two positive solutions, one positive solution, and no positive solution
for $b > b_{c}$, $b = b_{c}$ and $b < b_{c}$, respectively,
where 
\begin{eqnarray}\label{eq:b_critical}
b_{c} 
\equiv
\left( \frac{n+2}{n} \right)^{\frac{1}{2}} \left( \frac{n+2}{2} \right)^{\frac{1}{n}} r_{g},
\end{eqnarray}
is the critical impact parameter.
From Eqs. (\ref{eq:Photon_trajectory}) and (\ref{eq:G_definition}), 
we can see that the photon is scattered if $b > b_{c}$ while it reaches the event horizon $r=r_{g}$ if $b < b_{c}$. 

We will assume $b_{c}<b$ in what follows since we are interested in the scattering problem.
In this case, the larger positive solution $r_{0}$ of the equation $G(r,b)=0$ is the closest distance of the photon.
From $G(r_{0},b)=0$, the relation between the impact parameter $b$ and the closest distance $r_{0}$ is given by
\begin{eqnarray}\label{eq:impact_closest_relation}
\frac{1}{b^{2}}=\frac{1}{r_{0}^{2}} \left[ 1- \left( \frac{r_{g}}{r_{0}} \right)^{n} \right].
\end{eqnarray}

The derivative of $G(r,b)$ with respect to $r$ is given by
\begin{eqnarray}
\frac{\partial G(r,b)}{\partial r} = \frac{2}{r^{3}} -(n+2) \frac{r_{g}^{n}}{r^{n+3}}.
\end{eqnarray}
From $\partial G(r,b)/\partial r=0$, the radius of the photon sphere $r_{m}$ is obtained as
\begin{eqnarray}\label{eq:photon_sphere}
r_{m} = \left( \frac{n+2}{2} \right)^{\frac{1}{n}}r_{g}.
\end{eqnarray}

The deflection angle $\alpha$ is given by
\begin{equation}\label{eq:deflection_angle} 
\alpha=I(b)-\pi ,
\end{equation}
where
\begin{equation} 
I(b) \equiv 2 \int^{\infty}_{r_{0}}\frac{dr}{r^{2} \sqrt{G(r,b)}}.
\end{equation}

\section{Deflection angle of light in weak field approximation}
In this section, we will calculate the deflection angle in the Tangherlini spacetime
in weak field approximation by Keeton and Petters' method \cite{Keeton_Petters_2006}.
Under the weak field approximation, the closest distance $r_{0}$ is much bigger than the radius of the photon sphere $r_{m}$.
We introduce a small parameter $h$ which is defined by
\begin{eqnarray}
h \equiv \left( \frac{r_{g}}{r_{0}} \right)^{n} \ll \left( \frac{r_{g}}{r_{m}} \right)^{n}=\frac{2}{n+2}.
\end{eqnarray}
The relation between the impact parameter $b$ and the closest distance $r_{0}$ (\ref{eq:impact_closest_relation}) is expressed by
\begin{eqnarray}
\left( \frac{r_{0}}{b} \right)^{2} =1-h.
\end{eqnarray}
Thus, the small amount $h$ is expressed by
\begin{eqnarray}
h =\left( \frac{r_{g}}{b} \right)^{n} +O(h^{2}).
\end{eqnarray}

Using $x \equiv r_{0}/r$, the deflection angle $\alpha$ is rewritten as
\begin{equation} 
\alpha = 2 \int^{1}_{0} \frac{dx}{\sqrt{1-x^{2}} \sqrt{1-hf(x)}}-\pi,
\end{equation}
where
\begin{equation} 
f(x) 
\equiv \frac{1-x^{n+2}}{1-x^{2}}
=\frac{1+x+x^{2}+\cdots +x^{n+1}}{1+x}.
\end{equation}
The function $f(x)$ is monotonically increasing with respect to $x$
and changes from $1$ to $(n+2)/2$
as $x$ increases from $0$ to $1$.
Note that $hf(x)$ is much smaller than the unity 
since
$hf(x) \ll \left( r_{g}/r_{m} \right)^{n} f(1)  
=1.$

The Taylor series of $(1-hf(x))^{-\frac{1}{2}}$ with respect to 
$hf(x)$ is obtained as
\begin{equation} 
(1-hf(x))^{-\frac{1}{2}}
=1+\frac{1}{2}hf(x)+O(h^{2}).
\end{equation}
Therefore, the deflection angle is given by
\begin{equation} 
\alpha = 2 \int^{1}_{0} \frac{dx}{\sqrt{1-x^{2}}} +h\int^{1}_{0} \frac{1-x^{n+2}}{(1-x^{2})^{\frac{3}{2}}}dx -\pi +O(h^{2}).
\end{equation}
We can easily integrate the first term as 
\begin{equation} 
\int^{1}_{0} \frac{dx}{\sqrt{1-x^{2}}}
=\left[ \arcsin x \right]^{1}_{0}
=\frac{\pi}{2}.
\end{equation}
Thus, the deflection angle is rewritten as
\begin{eqnarray}\label{eq:alpha_weak}
\alpha 
&=& H_{n+2}h +O(h^{2})\nonumber\\
&=& H_{n+2} \left( \frac{r_{g}}{b} \right)^{n} +O\left( \left( \frac{r_{g}}{b} \right)^{2n}\right),
\end{eqnarray}
where
\begin{equation} 
H_{m} \equiv \int^{\frac{\pi}{2}}_{0} \frac{1-\sin^{m}k }{\cos^{2}k}dk,
\end{equation}
where $k \equiv \arcsin x$ and $m$ is a positive integer.

A recurrence formula is obtained as
\begin{equation}\label{eq:Recurrence}
H_{n+2}=H_{n}+B_{n},
\end{equation}
where $B_{n}$ is 
\begin{eqnarray}\label{eq:A_n}
B_{n}
&\equiv& \int^{\frac{\pi}{2}}_{0} \sin^{n}k dk
= \int^{\frac{\pi}{2}}_{0} \cos^{n}k dk \nonumber\\
&=&\frac{\sqrt{\pi}}{2} \frac{\Gamma \left( \frac{n+1}{2} \right)}{\Gamma \left( \frac{n+2}{2} \right)} \nonumber\\
&=&
\left\{
\begin{array}{ll}
\frac{(n-1)!!}{n!!}\frac{\pi}{2} &\quad \mathrm{for\:  an\: even\:} n,  \\
\frac{(n-1)!!}{n!!} &\quad \mathrm{for\: an\: odd\:} n.
\end{array}
\right.
\end{eqnarray}

When $n$ is even, we can put $n=2L$, where $L$ is a positive integer.
From $H_{2}=\pi/2$ and Eqs. (\ref{eq:Recurrence}) and (\ref{eq:A_n}), we obtain
\begin{eqnarray} 
H_{n+2}
=H_{2}+\sum^{L}_{m=1}B_{2m}
=\frac{\pi}{2}\left[ 1+\sum^{L}_{m=1}\frac{(2m-1)!!}{(2m)!!} \right]. \qquad 
\end{eqnarray}
Thus, the deflection angle is obtained as
\begin{equation}\label{eq:deflection_angle_weak_even}
\alpha = \frac{\pi}{2}\left[ 1+\sum^{L}_{m=1}\frac{(2m-1)!!}{(2m)!!} \right] \left( \frac{r_{g}}{b} \right)^{n} +O\left( \left( \frac{r_{g}}{b} \right)^{2n}\right).
\end{equation}


When $n$ is odd, we can put $n=2L-1$.
From $H_{1}=1$ and Eqs. (\ref{eq:Recurrence}) and (\ref{eq:A_n}), we get
\begin{eqnarray} 
H_{n+2}
=H_{1}+\sum^{L}_{m=1}B_{2m-1}
=1+\sum^{L}_{m=1}\frac{(2m-2)!!}{(2m-1)!!}. \qquad 
\end{eqnarray}
Here we have defined $0!!=1$.
Thus, the deflection angle is given by
\begin{equation}\label{eq:deflection_angle_weak_odd}
\alpha = \left[ 1+\sum^{L}_{m=1}\frac{(2m-2)!!}{(2m-1)!!} \right] \left( \frac{r_{g}}{b} \right)^{n} +O\left( \left( \frac{r_{g}}{b} \right)^{2n}\right).
\end{equation}


We can also calculate the deflection angle by the nonlinear terms 
with respect to $h$ by Keeton and Petters' method \cite{Keeton_Petters_2006}.  
Our purpose in this paper is to research the relativistic images by exotic lens objects and the effect on the total magnification. 
Note that the first order term with respect to h in the weak gravitational field is enough for reaching our purpose.

\section{Deflection angle in strong field limit}
In this section, we will investigate the deflection angle in the Tangherlini spacetime in the strong field limit.
We will express the deflection angle $\alpha$ in the strong field limit by 
\begin{eqnarray}\label{eq:deflection_angle_strong_b}
\alpha(b)
=-\bar{a}\log \left( \frac{b}{b_{c}}-1 \right) + \bar{b} +O\left( (b-b_{c})^{\frac{1}{2}} \right),
\end{eqnarray}
or
\begin{eqnarray}\label{eq:deflection_angle_strong_theta}
\alpha(\theta)
=-\bar{a}\log \left( \frac{\theta D_{l}}{b_{c}}-1 \right) + \bar{b} +O\left( (\theta D_{l}-b_{c})^{\frac{1}{2}} \right),
\end{eqnarray}
where $\bar{a}$ is a positive parameter, $\bar{b}$ is a parameter, $\theta$ is the image angle 
and $D_{l}$ is the angular diameter distance between the observer and the lens object.
For a small image angle $\theta \ll 1$, the impact parameter $b$ is given by~\cite{Schneider_Ehlers_Falco_1992}
\begin{eqnarray}\label{eq:small_theta}
b=\theta D_{l}.
\end{eqnarray}
If we get the explicit expression for the deflection angle in the strong field limit, 
we can calculate a countably infinite number of relativistic image angles denoted by $\theta_{N}$ 
and the corresponding magnifications $\mu_{N}$ individually~\cite{Bozza_2002}.

We show the explicit expression for the divergent part of the deflection angle 
in the all-dimensional Tangherlini spacetime 
and we integrate the regular part of the deflection angle in $4$, $5$ and $7$ dimension.\footnote{As below, 
we obey the convention of the analysis in the strong field limit 
but the definitions of some symbols such as $z$ are different from the definitions by Bozza~\cite{Bozza_2002}.}
Using by Eq. (\ref{eq:impact_closest_relation}) and 
\begin{eqnarray}
z \equiv 1 -\left( \frac{r_{0}}{r} \right)^{n},
\end{eqnarray}
we rewrite $G(r,b)$ and $I(b)$ into  
\begin{eqnarray}
G(z,r_{0})
&=&\frac{1}{r_{0}^{2}} \left\{  1- \left( \frac{r_{g}}{r_{0}} \right)^{n} \right. \nonumber\\
&&\left.+(1-z)^{\frac{2}{n}} \left[  -1 +\left( \frac{r_{g}}{r_{0}} \right)^{n}(1-z) \right] \right\}
\end{eqnarray}
and
\begin{eqnarray}
I(r_{0})=\int^{1}_{0}R(z)f(z,r_{0})dz,
\end{eqnarray}
respectively, where
\begin{eqnarray}
R(z) \equiv \frac{2}{n}(1-z)^{\frac{1}{n}-1}
\end{eqnarray}
and
\begin{eqnarray}
f(z,r_{0}) 
&\equiv& \frac{1}{\sqrt{r_{0}^{2}G(z,r_{0})}} \nonumber\\
&=& \frac{1}{\sqrt{1- \left( \frac{r_{g}}{r_{0}} \right)^{n} 
+(1-z)^{\frac{2}{n}} \left[ -1 +\left( \frac{r_{g}}{r_{0}} \right)^{n}(1-z) \right] }}.\nonumber\\
\end{eqnarray}


We expand $r_{0}^{2}G(z,r_{0})$ near $z=0$ and obtain
\begin{eqnarray}
r_{0}^{2}G(z,r_{0})
=\gamma(r_{0})z+\beta(r_{0})z^{2}+\cdots ,
\end{eqnarray}
where
\begin{eqnarray}
&&\gamma(r_{0}) \equiv \frac{1}{n} \left[ 2-(n+2) \left( \frac{r_{g}}{r_{0}} \right)^{n} \right] \\
&&\beta(r_{0}) \equiv \frac{1}{n^{2}} \left[ n -2 +(n+2) \left( \frac{r_{g}}{r_{0}} \right)^{n} \right].
\end{eqnarray}
Near the photon sphere $r_{0}=r_{m}$, $\gamma(r_{0})$ and $\beta(r_{0})$ are expanded as
\begin{eqnarray}
\gamma(r_{0}) = \frac{2}{r_{m}}(r_{0}-r_{m}) +O\left( (r_{0}-r_{m})^{2} \right)
\end{eqnarray}
and
\begin{eqnarray}
\beta(r_{0}) =\frac{1}{n} -\frac{2}{nr_{m}}(r_{0}-r_{m}) +O\left( (r_{0}-r_{m})^{2} \right).
\end{eqnarray}

We will divide $I(r_{0})$ into the divergent part $I_{D}(r_{0})$ and the regular part $I_{R}(r_{0})$ as
\begin{equation}
I(r_{0})=I_{D}(r_{0})+I_{R}(r_{0}).
\end{equation}
The divergent part $I_{D}(r_{0})$ is defined as
\begin{equation}
I_{D}(r_{0}) \equiv \int^{1}_{0} R(0)f_{0}(z,r_{0})dz,
\end{equation}
where
\begin{eqnarray}
f_{0}(z,r_{0}) 
\equiv \frac{1}{\sqrt{\gamma(r_{0})z+\beta(r_{0})z^{2}}}.
\end{eqnarray}
By a straightforward calculation, the divergent part $I_{D}(r_{0})$ is obtained as
\begin{eqnarray}
&&I_{D}(r_{0}) \nonumber\\
&&=\frac{2}{n\sqrt{\beta(r_{0})}} \log \left| \frac{\gamma(r_{0})+2\beta(r_{0})+2\sqrt{(\gamma(r_{0})+\beta(r_{0}))\beta(r_{0})}}{\gamma(r_{0})} \right| \nonumber\\ 
&&=\frac{4}{n\sqrt{\beta(r_{0})}} \log \left( \frac{\sqrt{\beta(r_{0})}+\sqrt{\gamma(r_{0})+\beta(r_{0})}}{\sqrt{\gamma(r_{0})}} \right).
\end{eqnarray}
Therefore, the divergent part $I_{D}(r_{0})$ is expressed by
\begin{eqnarray}
I_{D}(r_{0})
= -\frac{2}{\sqrt{n}}\log \left( \frac{r_{0}}{r_{m}}-1 \right) +\frac{2}{\sqrt{n}}\log \frac{2}{n} +O(r_{0}-r_{m}).\nonumber\\
\end{eqnarray}

We will rewrite the divergent part $I_{D}(r_{0})$ into a function $I_{D}(b)$ with respect to the impact parameter $b$ 
since the lens equation is usually written as an equation in terms of the impact parameter $b$ or the image angle $\theta$. 
From the relation between the impact parameter~$b$ and the closest distance~$r_{0}$~(\ref{eq:impact_closest_relation}), 
we can regard the impact parameter $b(r_{0})$ as a function of the closest distance $r_{0}$.
Using by Eq.~(\ref{eq:impact_closest_relation}), we expand the impact parameter $b(r_{0})$ in a series near $r_{0}=r_{m}$ to get
\begin{eqnarray}\label{eq:b_r0}
b(r_{0})
&=&b_{c} + \frac{1}{2} \left( \frac{n+2}{n} \right)^{\frac{3}{2}}\frac{n}{r_{m}}(r_{0}-r_{m})^{2} \nonumber\\
&&+O\left((r_{0}-r_{m})^{3}\right). 
\end{eqnarray}
From Eqs. (\ref{eq:b_critical}), (\ref{eq:photon_sphere}) and (\ref{eq:b_r0}), we obtain
\begin{eqnarray}
\log \left( \frac{r_{0}}{r_{m}}-1 \right)
&=&\frac{1}{2} \log \left( \frac{b}{b_{c}}-1 \right) +\frac{1}{2}\log \left( \frac{2}{n+2} \right) \nonumber\\
&&+O(r_{0}-r_{m}).
\end{eqnarray}
Hence, the divergent part is rewritten as
\begin{eqnarray}
I_{D}(b)
&=&-\frac{1}{\sqrt{n}}\log \left( \frac{b}{b_{c}}-1 \right)+\frac{1}{\sqrt{n}}\log \frac{2(n+2)}{n^{2}}\nonumber\\
&&+O\left( (b-b_{c})^{\frac{1}{2}} \right).
\end{eqnarray}

The regular part $I_{R}(r_{0})$ is defined as
\begin{eqnarray}
I_{R}(r_{0}) \equiv \int^{1}_{0}g(z,r_{0})dz,
\end{eqnarray}
where 
\begin{eqnarray}
g(z,r_{0}) \equiv R(z)f(z,r_{0})- R(0)f_{0}(z,r_{0}).
\end{eqnarray}
We expand $I_{R}(r_{0})$ in powers of $(r_{0}-r_{m})$ 
and express it as a function $I_{R}(b)$ with respect to $b$ in the following form:
\begin{eqnarray}
I_{R}(r_{0})
& 
=
& 
\sum^{\infty}_{l=0}\frac{1}{l!}(r_{0}-r_{m})^{l}\int^{1}_{0} \left. \frac{\partial^{l}g}{\partial r_{0}^{l}}\right|_{r_{0}=r_{m}}dz
\nonumber\\ 
&=&\frac{2}{n} \int^{1}_{0} \left[ \frac{\sqrt{n+2}(1-z)^{\frac{1}{n}-1}}{\sqrt{n-(1-z)^\frac{2}{n}(n+2z)}} -\frac{\sqrt{n}}{z} \right]dz \nonumber\\
&&+O(r_{0}-r_{m}) \nonumber\\
&=&2\sqrt{n+2}\int^{1}_{0}\frac{dy}{\sqrt{n-(n+2)y^{2}+2y^{n+2}}}\nonumber\\
&&-\frac{2\sqrt{n}}{n}\int^{1}_{0}\frac{dz}{z} +O\left( (b-b_{c})^{\frac{1}{2}} \right) \nonumber\\
&=&I_{R}(b),
\end{eqnarray}
where we have used $y \equiv (1-z)^{\frac{1}{n}}$.

Thus, the deflection angle $\alpha (b)$ in the strong field limit is obtained as
\begin{eqnarray}
\alpha(b)
&=&I_{D}(b)+I_{R}(b)-\pi \nonumber\\
&=&-\frac{1}{\sqrt{n}}\log \left( \frac{b}{b_{c}}-1 \right)+\frac{1}{\sqrt{n}}\log \frac{2(n+2)}{n^{2}} \nonumber\\
&&+I_{R}(b) -\pi +O\left( (b-b_{c})^{\frac{1}{2}} \right).
\end{eqnarray}
Hence, we get the parameters $\bar{a}=\frac{1}{\sqrt{n}}$ and $\bar{b}=\frac{1}{\sqrt{n}}\log \frac{2(n+2)}{n^{2}}+I_{R}(b) -\pi$.

We can analytically calculate the regular parts $I_{R}(b)$ for $n=1$, $2$ and $4$ since the elliptic functions $I(b)$ for $n=1$, $2$ and $4$ are integrable~\cite{Gibbons_Vyska_2012}.

\subsection{$n=1$}
We consider the case for $n=1$.
In this case, the critical impact parameter and the radius of the photon sphere 
are given by $b_{c}= \frac{3\sqrt{3}r_{g}}{2}$ and $r_{m} = \frac{3r_{g}}{2}$, respectively.
The divergent part and the regular part of the deflection angle are obtained as
\begin{eqnarray}
I_{D}(b)
=-\log \left( \frac{b}{b_{c}}-1 \right)+\log 6 +O\left( (b-b_{c})^{\frac{1}{2}} \right)
\end{eqnarray}
and
\begin{eqnarray}
I_{R}(b)
&=&2\int^{1}_{0} \left( \frac{1}{z\sqrt{1-\frac{2}{3}z}} -\frac{1}{z}  \right) dz +O\left( (b-b_{c})^{\frac{1}{2}} \right) \nonumber\\
&=&2\log \left[ 6 \left( 2-\sqrt{3} \right) \right] +O\left( (b-b_{c})^{\frac{1}{2}} \right),
\end{eqnarray}
respectively.
Thus, the deflection angle $\alpha(b)$ is obtained as 
\begin{eqnarray}\label{eq:deflection_angle_strong_n=1}
\alpha(b)
&=&I_{D}(b)+I_{R}(b)-\pi \nonumber\\
&=&-\log \left( \frac{b}{b_{c}}-1 \right) 
+ \log \left[ 216 \left( 7-4\sqrt{3} \right) \right] \nonumber\\ 
&&-\pi +O\left( (b-b_{c})^{\frac{1}{2}} \right).
\end{eqnarray}
Therefore, we get the parameters $\bar{a}=1$ and $\bar{b}=\log \left[ 216 \left( 7-4\sqrt{3} \right) \right] \ -\pi \simeq -0.40$.
It recovers the deflection angle of the light in the Schwarzschild spacetime in the strong field limit 
which was obtained by Bozza~\cite{Bozza_2002}.

\subsection{$n=2$}
For $n=2$, the critical impact parameter and the radius of the photon sphere 
are $b_{c}= 2r_{g}$ and $r_{m} = \sqrt{2}r_{g}$, respectively.
The divergent part and the regular part of the deflection angle are obtained as
\begin{eqnarray}
I_{D}(b)
&=&-\frac{1}{\sqrt{2}} \log \left( \frac{b}{b_{c}}-1 \right)+ \frac{1}{\sqrt{2}}\log 2 \nonumber\\
&&+O\left( (b-b_{c})^{\frac{1}{2}} \right).
\end{eqnarray}
and
\begin{eqnarray}
I_{R}(b)
&=&\int^{1}_{0} \left( \frac{\sqrt{2}}{z\sqrt{1-z}} -\frac{\sqrt{2}}{z} \right)dz +O\left( (b-b_{c})^{\frac{1}{2}} \right) \nonumber\\
&=&2\sqrt{2}\log 2 +O\left( (b-b_{c})^{\frac{1}{2}} \right),
\end{eqnarray}
respectively.
Thus, the deflection angle $\alpha(b)$ is obtained as
\begin{eqnarray}\label{eq:deflection_angle_strong_n=2}
\alpha(b)
&=&-\frac{1}{\sqrt{2}} \log \left( \frac{b}{b_{c}}-1 \right)+ \frac{5\sqrt{2}}{2}\log 2 -\pi \nonumber\\
&&+O\left( (b-b_{c})^{\frac{1}{2}} \right).
\end{eqnarray}
In this case, the parameters are obtained as $\bar{a}=\frac{1}{\sqrt{2}}$ and $\bar{b}=\frac{5\sqrt{2}}{2}\log 2 -\pi \sim -0.69$.

\subsection{$n=4$}
For $n=4$, the critical impact parameter and the radius of the photon sphere are given by 
$b_{c}=\left( \frac{27}{4} \right)^{\frac{1}{4}}r_{g}$ and $r_{m}=3^{\frac{1}{4}}r_{g}$, respectively.
The divergent part and the regular part of the deflection angle are obtained as
\begin{eqnarray}
I_{D}(b)
&=&-\frac{1}{2} \log \left( \frac{b}{b_{c}}-1 \right)+ \frac{1}{2}\log \frac{3}{4} \nonumber\\
&&+O\left( (b-b_{c})^{\frac{1}{2}} \right)
\end{eqnarray}
and
\begin{eqnarray}
I_{R}(b)
&=&2\sqrt{3}\int^{1}_{0}\frac{dy}{\sqrt{2-3y^{2}+y^{6}}}-\int^{1}_{0}\frac{dz}{z} +O\left( (b-b_{c})^{\frac{1}{2}} \right) \nonumber\\
&=&\log 12 +O\left( (b-b_{c})^{\frac{1}{2}} \right),
\end{eqnarray}
respectively. The deflection angle is obtained as 
\begin{eqnarray}\label{eq:deflection_angle_strong_n=4}
\alpha(b)
=
-\frac{1}{2}\log \left( \frac{b}{b_{c}}-1 \right)+\log 6\sqrt{3} -\pi +O\left( (b-b_{c})^{\frac{1}{2}} \right) 
\nonumber\\
\end{eqnarray}
and hence we get the parameter $\bar{a}=\frac{1}{2}$ and $\bar{b}=\log 6\sqrt{3} -\pi \sim -0.80$.

In Fig.~1, we have plotted the deflection angles for $n=1$, $2$ and $4$ 
to verify the accuracy of the approximations in the weak and strong gravitational fields. 
We can see that the approximation error in the weak field depends on the value of $n$. 
\begin{figure}[htbp]
\begin{center}
\includegraphics[width=79mm]{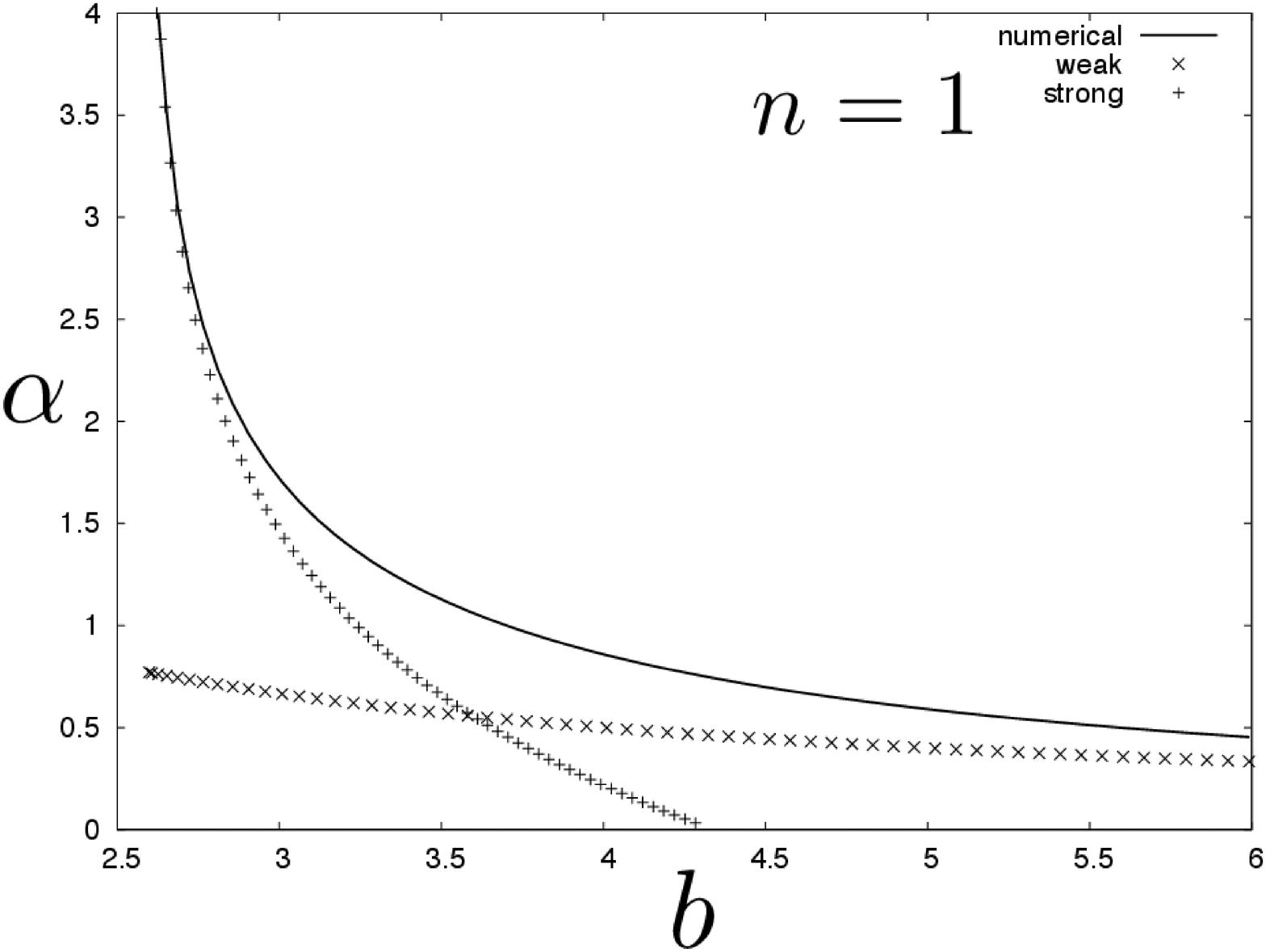}
\includegraphics[width=79mm]{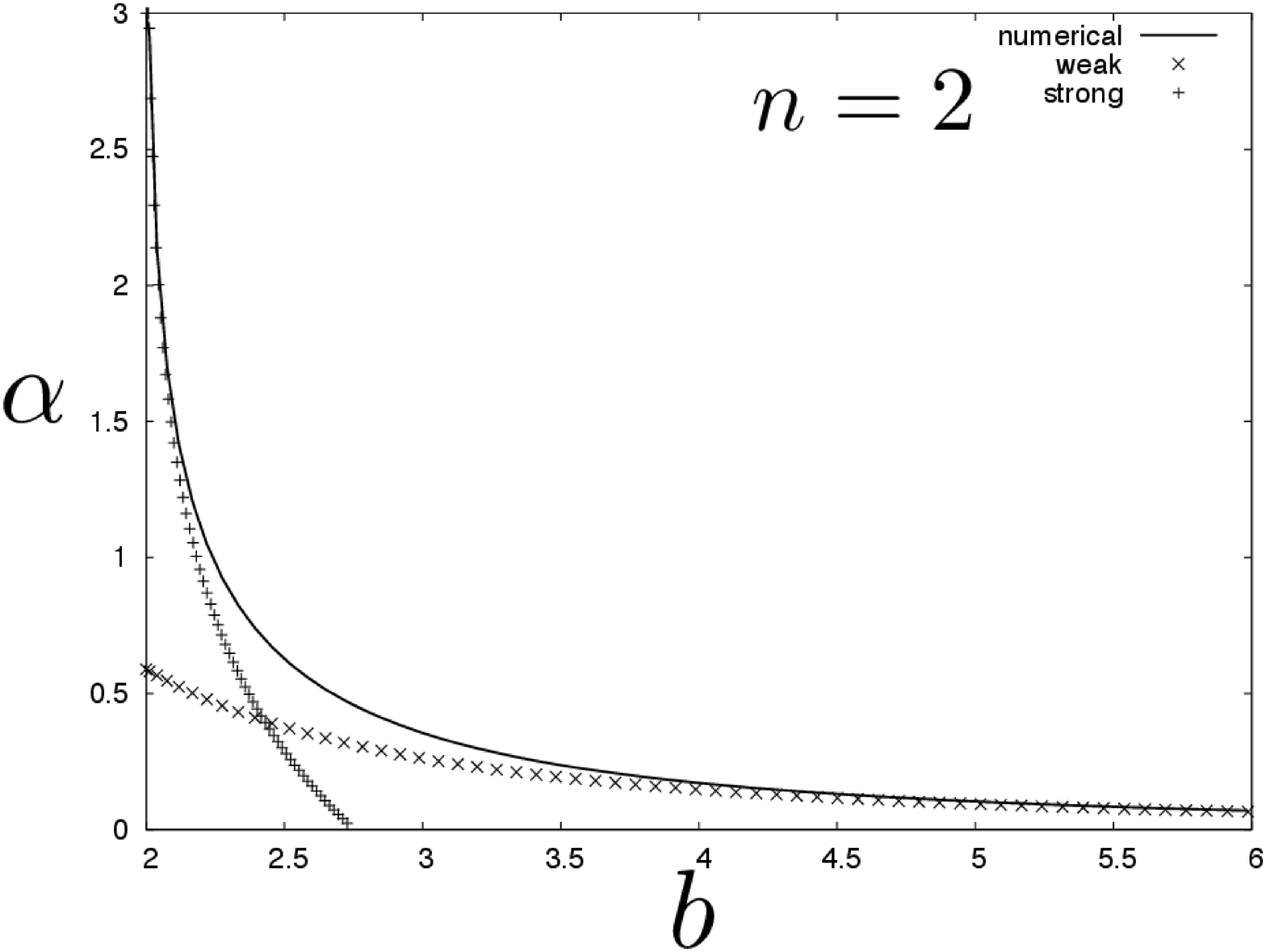}
\includegraphics[width=79mm]{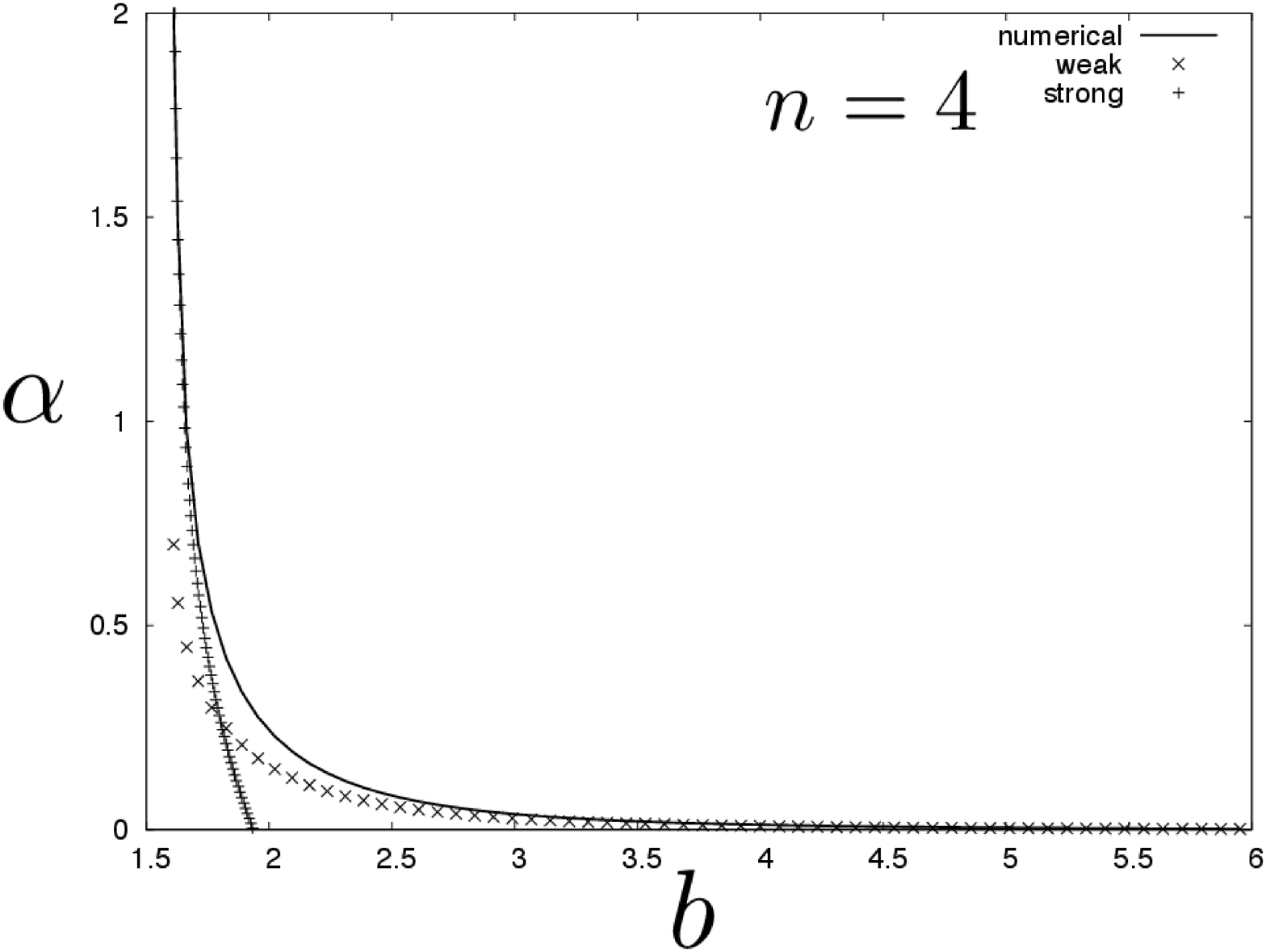}
\end{center}
\caption{Bending angle of light. Top panel (n=1), middle (n=2), and bottom (n=4).
 The horizontal axis denotes $b$ in the unit of $r_g$.
 The solid curve, $+$, and $\times$ correspond to numerical calculations,
 the strong-field limit, and the weak-field approximation, respectively.
}
\end{figure}

\section{Gravitational Lensing}
\subsection{Lens equation}
We consider the lens configuration which is shown by Fig.~2. 
The light ray which is emitted by the source $S$ bends near the lensing object $L$. 
The observer $O$ does not see the source $S$ with the source angle $\phi$ but the image $I$ with the image angle $\theta$.
For simplicity, we assume that 
both the observer $O$ and the source $S$ are far from the lensing object $L$ or $D_{l}\gg b$ and $D_{ls}\gg b$, 
where $D_{l}$ and $D_{ls}$ are the angular diameter distances~\cite{Schneider_Ehlers_Falco_1992} between $O$ and $L$ 
and between $L$ and $S$.
We also assume the thin lens approximation that the light ray bends on the lens plane.
The impact parameter $b$ is described by $b=D_{l}\theta$.
Under the assumptions, the effective deflection angle $\bar{\alpha}$, the source angle $\phi$ and the image angle $\theta$ are small 
or $|\bar{\alpha}|\ll 1$, $|\phi|\ll 1$ and $|\theta|\ll 1$.
The effective deflection angle $\bar{\alpha}$ is defined by 
\begin{eqnarray}
\bar{\alpha} \equiv ( \alpha \;  \mathrm{mod} \; 2\pi ).
\end{eqnarray}
The deflection angle $\alpha$ is expressed by
\begin{eqnarray}\label{eq:alpah_baralpha}
\alpha =\bar{\alpha} +2\pi N,
\end{eqnarray}
where $N$ is a non-negative integer which denotes the winding number of the light ray.

\begin{figure}[htbp]
\begin{center}
\includegraphics[width=80mm]{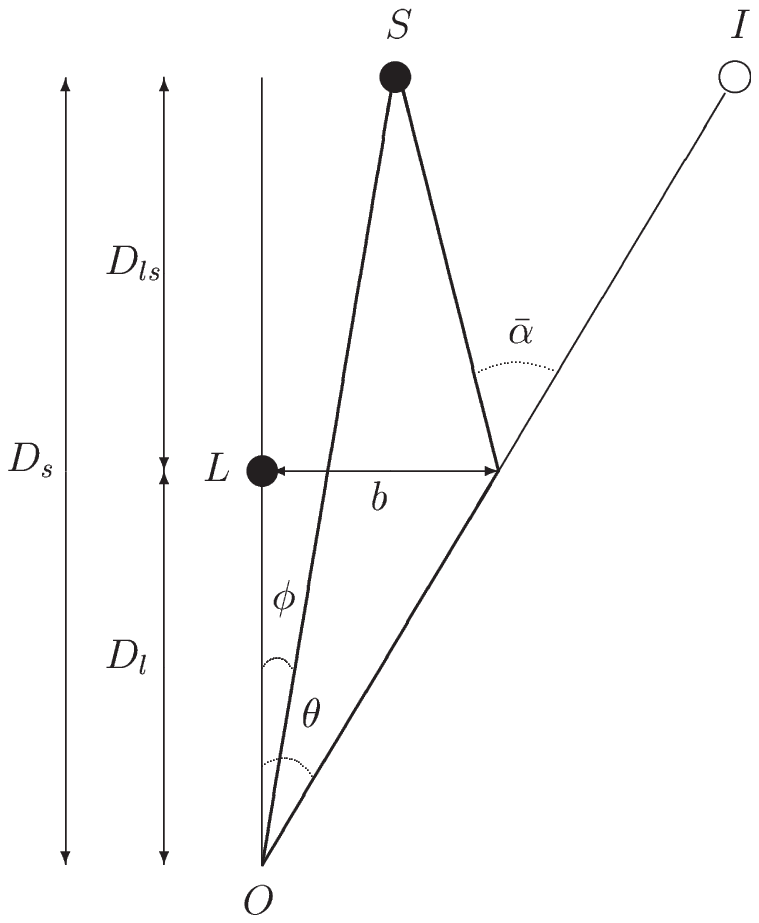}
\end{center}
\caption{The configuration of the gravitational lensing. The light ray emitted by the source object $S$ bends on the lens plane. 
The observer $O$ does not see the source $S$ with the source angle $\phi$ but the image $I$ with the image angle $\theta$.
$\bar{\alpha}$ is the effective deflection angle and $b$ is the impact parameter of the light.
$D_{l}$ and $D_{ls}$ are the angular diameter distances between the observer $O$ and the lens object $L$ 
and between the lens object $L$ and the source object $S$, respectively.
The angular diameter distance between the observer $O$ and the source $S$ is given by $D_{s}=D_{l}+D_{ls}$.  
}
\end{figure}

Then, the lens equation is given by~\cite{Tsukamoto_Harada_Yajima_2012}
\begin{eqnarray}\label{eq:Lens_equation}
D_{ls}\bar{\alpha}=D_{s}(\theta-\phi),
\end{eqnarray}
where $D_{s}$ is the angular diameter distance between the observer $O$ and the source $S$ and satisfies the relation $D_{s}=D_{l}+D_{ls}$.
In the case $N=0$, the lens equation becomes the standard one~\cite{Schneider_Ehlers_Falco_1992} because of $\bar{\alpha}=\alpha$.
Note that there might exist small correction terms depending on formulations of the lens equation (e.g.~\cite{Virbhadra_Ellis_2000}). 
If the source angle $\phi=0$, ring-shaped images which are called the Einstein ring with the angle $\theta_{0}$ for $N=0$ 
and the relativistic Einstein ring with the angle $\theta_{N\geq 1}$ for $N\geq 1$ appear from the symmetry. 
From $N=0$, $\phi=0$ and Eqs.  (\ref{eq:alpha_weak}), (\ref{eq:small_theta}), (\ref{eq:alpah_baralpha}) and (\ref{eq:Lens_equation}), 
the Einstein ring angle is obtained as
\begin{eqnarray}
\theta_{0}
\sim \left( H_{n+2}\frac{D_{ls}}{D_{s}} \right)^{\frac{1}{n+1}} \left( \frac{r_{g}}{D_{l}} \right)^{\frac{n}{n+1}}.
\end{eqnarray}

The behaviors of the Tangherlini lens model in the weak field approximation 
have been known already because it is included in an exotic lens model 
or a general spherical lens model~\cite{Kitamura_Nakajima_Asada_2013,Tsukamoto_Harada_2013,
Izumi_Hagiwara_Nakajima_Kitamura_Asada_2013,Kitamura_Izumi_Nakajima_Hagiwara_Asada_2013}.
Here, we refer only to the image angles and the magnification in the directly aligned limit ($|\phi| \ll \theta_{0} \ll 1$).
Under the weak approximation, the lens equation has two 
solutions~$\theta_{0+}$ and $\theta_{0-}$ regardless of the source angle $\phi$ for $n\geq 1$.
For $|\phi| \ll \theta_{0} \ll 1$,
the image angles $\theta_{0\pm}$ and the magnification $\mu_{0\pm}$ are obtained as~\cite{Kitamura_Nakajima_Asada_2013,Tsukamoto_Harada_2013}
\begin{eqnarray}
\theta_{0\pm}=\pm \theta_{0} + \frac{\phi}{1+n} \pm \frac{n\phi^{2}}{2(1+n)^{2}\theta_{0}} +O\left(\frac{\phi^{3}}{\theta_{0}^{2}}\right)
\end{eqnarray}
and
\begin{eqnarray}
\mu_{0\pm} \sim \frac{1}{1+n}\frac{\phi \pm \theta_{0}}{\phi},
\end{eqnarray}
respectively.
The total magnification $\mu_{0}$ in the directly aligned limit is obtained as
\begin{eqnarray}
\mu_{0}\equiv \left|\mu_{0+}\right|+\left|\mu_{0-}\right| = \frac{2}{1+n}\frac{\theta_{0}}{\phi}.
\end{eqnarray}
Here, we assume that light rays (as electromagnetic waves) travel
only on our four-dimensional slice in Tangherlini solution.
Therefore, light rays that travel through the additional dimensions
are ignored in the calculation of the magnification.

The relativistic Einstein rings or the relativistic images always appear on the region just outside the photon sphere.
The angle of the innermost relativistic Einstein ring is obtained as
\begin{eqnarray}
\theta_{\infty}
=\frac{b_{c}}{D_{l}}
=\left( 1+ \frac{2}{n} \right)^{\frac{1}{2}} \left( 1+\frac{n}{2} \right)^{\frac{1}{n}} \frac{r_{g}}{D_{l}}.
\end{eqnarray}
The relation between the Einstein ring angle $\theta_{0}$, the relativistic Einstein ring angle $\theta_{\infty}$ 
and the relativistic image angle $\theta_{N \geq 1}(\phi)$ is obtained as
\begin{eqnarray}
\theta_{N \geq 1}(\phi)
&\sim& \theta_{\infty} \nonumber\\
&\sim& \sqrt{\frac{n+2}{n}} \left( \frac{n+2}{2H_{n+2}}\frac{D_{s}}{D_{ls}} \right)^{\frac{1}{n}} \theta_{0}^{\frac{n+1}{n}}.
\end{eqnarray}

\subsection{Magnifications and angles of the relativistic images}
We will investigate the magnifications $\mu_{N \geq 1}$ 
and the angles $\theta_{N \geq 1}$ of the relativistic images~\cite{Bozza_Capozziello_Iovane_Scarpetta_2001,Bozza_2002} 
in the Tangherlini spacetime.
We use the deflection angle $\alpha(\theta)$ (\ref{eq:deflection_angle_strong_theta}) in this subsection.

For the winding number $N\geq 1$, we define an angle $\theta_{N\geq 1}^0$ as
\begin{eqnarray}\label{eq:C5_theta}
\alpha(\theta_{N \geq 1}^0)=2\pi N.
\end{eqnarray}
From Eqs. (\ref{eq:deflection_angle_strong_theta}) and (\ref{eq:C5_theta}), we obtain 
\begin{eqnarray}\label{eq:theta_0_n}
\theta_{N \geq 1}^0
=\frac{b_{c}}{D_{l}} \left[ 1+e^{ \left( \bar{b}-2\pi N \right) \sqrt{n} }  \right].
\end{eqnarray}
We expand the deflection angle $\alpha (\theta)$ around $\theta=\theta^{0}_{N \geq 1}$ to obtain the effective deflection angle $\bar{\alpha}$.
We define a small angle 
\begin{eqnarray}\label{eq:small_image_angle}
\Delta \theta_{N \geq 1} \equiv \theta_{N \geq 1}(\phi) - \theta_{N \geq 1}^{0},
\end{eqnarray}
where $\theta_{N \geq 1}(\phi)$ is the solution of the lens equation~(\ref{eq:Lens_equation}) for a winding number $N \geq 1$,
namely the relativistic image angle.
From Eqs. (\ref{eq:deflection_angle_strong_theta}) and (\ref{eq:theta_0_n}), 
the effective deflection angle in the strong field limit is obtained as
\begin{eqnarray}\label{eq:alpha_bar}
\bar{\alpha}=-\frac{D_{l}}{b_{c}}\frac{e^{\sqrt{n} \left(-\bar{b}+2\pi N \right) }}{\sqrt{n}}   \Delta \theta_{N \geq 1}.
\end{eqnarray}

We substitute the effective deflection angle (\ref{eq:alpha_bar}) into the lens equation~(\ref{eq:Lens_equation}) {\ and we obtain} 
\begin{eqnarray}\label{eq:phi_small}
\phi=\theta_{N \geq 1}^{0} + \left[ 1+ \frac{D_{l}}{b_{c}} \frac{D_{ls}}{D_{s}} \frac{e^{ \sqrt{n} \left(-\bar{b}+2\pi N \right) }}{\sqrt{n}}  \right] \Delta \theta_{N \geq 1}. \nonumber\\
\end{eqnarray}
From Eqs. (\ref{eq:small_image_angle}) and (\ref{eq:phi_small}), the relativistic image angle $\theta_{N\geq 1}(\phi)$ is obtained as
\begin{eqnarray}\label{eq:theta_n_bigger_1_resolt}
\theta_{N \geq 1}(\phi)
\simeq \theta_{N \geq 1}^{0} + \frac{b_{c}}{D_{l}} \frac{D_{s}}{D_{ls}} \sqrt{n} e^{ \sqrt{n}(\bar{b}-2\pi N)}  \left( \phi-\theta^{0}_{N \geq 1} \right),\nonumber\\
\end{eqnarray}
where we have used $b_{c}/D_{l} \ll 1$.
Equation~(\ref{eq:theta_n_bigger_1_resolt}) is valid for the relativistic images not only on the same side of the source 
but also on the opposite side. 
The relativistic image angle on the opposite side is obtained as $\theta_{N \geq 1}(-\phi)$~\cite{Bozza_2002,Virbhadra_Ellis_2000}. 
From Eqs. (\ref{eq:theta_0_n}) and (\ref{eq:theta_n_bigger_1_resolt}), the innermost relativistic image angle is obtained as
\begin{eqnarray}\label{eq:theta_infinity}
\theta_{\infty }=\theta^{0}_{\infty}=\frac{b_{c}}{D_{l}}.
\end{eqnarray}
From Eqs. (\ref{eq:theta_0_n}), (\ref{eq:theta_n_bigger_1_resolt}) and (\ref{eq:theta_infinity}), 
the difference of the angles between the outermost relativistic image and innermost one is given by
\begin{eqnarray}\label{eq:C5_difference}
\theta_{1}-\theta_{\infty}
\simeq \theta^{0}_{1}-\theta_{\infty}
=\theta_{\infty}e^{\sqrt{n}(\bar{b}-2\pi)}.
\end{eqnarray}

The magnification $\mu_{N \geq 1}$ of the relativistic image is obtained as
\begin{eqnarray}\label{eq:mu_n}
\mu_{N \geq 1}
&\simeq& \frac{\theta_{N \geq 1}}{\phi}\left. \frac{d\theta_{N \geq 1}}{d \phi} \right|_{\theta_{N \geq 1}=\theta_{N \geq 1}^{0}} \nonumber\\
&\simeq& \frac{1}{\phi}\frac{b_{c}^2}{D_{l}^2}\frac{D_{s}}{D_{ls}}\sqrt{n}e^{\sqrt{n}(\bar{b}-2\pi N)}.
\end{eqnarray}
The sum of the magnifications of all the relativistic images 
on one side of the source
is given by
\begin{eqnarray}
\sum^{\infty}_{N=1}\mu_{N}
&\simeq& \mu_{1}
\simeq \frac{1}{\phi}\frac{b_{c}^2}{D_{l}^2}\frac{D_{s}}{D_{ls}}\sqrt{n}e^{\sqrt{n}(\bar{b}-2\pi)} \nonumber\\
&\simeq& \frac{2}{H_{n+2}^{\frac{2}{n}} \sqrt{n} \phi} \left[ \frac{(n+2)D_s}{2D_{ls}} \right]^{\frac{2}{n}+1}
\theta_{0}^{\frac{2n+2}{n}} e^{\sqrt{n}(\bar{b}-2\pi)}. \nonumber\\
\end{eqnarray}
In the directly aligned limit, the ratio of the total magnification of the weak field 
images to the total magnification of all the relativistic images on both sides is obtained as
\begin{eqnarray}
\frac{\mu_{0}}{2\sum^{\infty}_{N=1} \left| \mu_{N} \right|}
&\simeq& \frac{\mu_{0}}{2\left| \mu_{1} \right|}  \nonumber\\
&\simeq&
\frac{H_{n+2}^{\frac{2}{n}} \sqrt{n}}{2(n+1)} \left[ \frac{2D_{ls}}{(n+2)D_{s}\theta_{0}} \right]^{\frac{2}{n}+1} e^{\sqrt{n}(2\pi-\bar{b})}. \nonumber\\
\end{eqnarray}
This shows that the relativistic images are always fainter than images in the weak field.

The total magnification of all the relativistic images 
on one side of the source can
become larger than $0.01$ when the source angle is 
\begin{eqnarray}
\left| \phi \right|
<
\frac{200}{H_{n+2}^{\frac{2}{n}} \sqrt{n}} \left[ \frac{(n+2)D_s}{2D_{ls}} \right]^{\frac{2}{n}+1} \theta_{0}^{\frac{2n+2}{n}} e^{\sqrt{n}(\bar{b}-2\pi)}. \qquad 
\end{eqnarray}
On the other hand, the demagnification of the images in the weak field could occur 
if $2\theta_{0}/(n+1)< \left| \phi \right|$ for $n>1$~\cite{Kitamura_Nakajima_Asada_2013}.  
The light curves calculated numerically have gutters of maximally $\sim 4$\%, $\sim 10$\% and $\sim 60$\% 
for $n=2$, $3$ and $10$, respectively, 
under the weak field approximation~\cite{Abe_2010,Kitamura_Nakajima_Asada_2013}.
These two regions for $\phi$ do not overlap because of $\theta_{0}^{(2n+2)/2} \ll \theta_{0}$.
Thus, the time-symmetric demagnification
of the light curve will appear even after taking account of the images in the strong gravitational field 
for $n>1$.

Before closing this section, let us illustrate an order-of-magnitude estimation
of the Tangherlini lens effects. 
Here, we assume a compact lens object which has a solar-mass-size photon sphere 
with $D_{l}=D_{ls}=10$ kpc and $r_{g}=3$ km in our Galaxy.
For $n=1$, $2$ and $4$, the angles of the Einstein ring 
$\theta_{0}$ are 
$3\times 10^{-9}$ rad, $5\times 10^{-12}$ rad and $3\times 10^{-14}$ rad,
the angles of the innermost relativistic Einstein ring $\theta_{\infty}$ are
$3\times 10^{-17}$ rad, $2\times 10^{-17}$ rad and $2\times 10^{-17}$ rad 
and the ratios of the total magnification $\mu_{0}/2\sum^{\infty}_{N=1} \left| \mu_{N} \right|$ in the directly aligned limit are 
$1\times 10^{27}$, $3\times 10^{25}$ and $8\times 10^{24}$, respectively.
Thus, we can neglect the relativistic images because they are too faint regardless of $n$.

\section{Summary}
We discussed whether relativistic images by the strong field can be neglected in general. 
It is likely that a spacetime geometry with $1/r^n$
fall-off in the weak field is not unique but has many variants.
The Tangherlini solution is one of the simplest models for our purpose.
For recent investigations of exotic gravitational lenses, therefore,
it would be of physical interest to examine Tangherlini lens both
in the weak field and in the strong one.

We calculated the divergent part of the deflection angle for arbitrary $n$
and the regular part for $n=1$, $2$ and $4$ in the strong field limit, 
the deflection angle for arbitrary $n$ under the weak gravitational approximation 
and the relation between the size of the Einstein ring and the ones of the relativistic Einstein rings for arbitrary $n$.
We showed that the relativistic images are always fainter than the images in the weak gravitational field.

We conclude that the images in the strong gravitational field have little effect on the total light curve
and that the time-symmetric demagnification~\cite{Abe_2010,Kitamura_Nakajima_Asada_2013} 
of the light curve will appear even after taking account of the images in the strong gravitational field 
for $n>1$.

\section*{Acknowledgements}
The authors would like to thank T. Harada and F. Abe for valuable comments and discussion.
N.T. is supported by the NSFC Grant No. 11305038, 
the Shanghai Municipal Education Commission grant for Innovative Programs No. 14ZZ001, 
the Thousand Young Talents Program, 
and Fudan University.


\end{document}